\documentclass[fleqn,usenatbib]{rasti}

\usepackage{newtxtext,newtxmath}

\usepackage[T1]{fontenc}

\usepackage{graphicx}	
\usepackage{amsmath}	
\usepackage[dvipsnames]{xcolor}
\usepackage{caption}
\usepackage{orcidlink}

\newcommand{\Rjup}{R$_{\mathrm{J}}$}
\newcommand{\Mjup}{M$_{\mathrm{J}}$}
\newcommand{\Rp}{R$_{\mathrm{P}}$}
\newcommand{\Mp}{M$_{\mathrm{P}}$}

\newcommand{\Teq}{T$_{\mathrm{eq}}$}
\newcommand{\degree}{$^{\circ}$}
\newcommand{\obliquity}{$|\lambda|$}
\defcitealias{Lustig-YaegerFu2023}{Lustig-Yaeger \& Fu et al.}
\defcitealias{PenzlinBooth2024}{Penzlin \& Booth et al.}

\title[BOWIE-ALIGN: the survey]{BOWIE-ALIGN: A JWST comparative survey of aligned vs misaligned hot Jupiters to test the dependence of atmospheric composition on migration history}

\author[J. Kirk et al.]{James Kirk$^{\orcidlink{0000-0002-4207-6615},1}$\thanks{E-mail: j.kirk22@imperial.ac.uk (JK)},
Eva-Maria Ahrer$^{\orcidlink{0000-0003-0973-8426},2}$,
Anna B.T. Penzlin$^{{\orcidlink{0000-0002-8873-6826}},1}$,
James E. Owen$^{\orcidlink{0000-0002-4856-7837},1}$, 
Richard A. Booth$^{{\orcidlink{0000-0002-0364-937X}},3}$, 
\newauthor
Lili Alderson$^{\orcidlink{0000-0001-8703-7751},4}$,
Duncan A. Christie$^{{\orcidlink{0000-0002-4997-0847}},2}$,
Alastair B. Claringbold$^{\orcidlink{0000-0003-1309-5558},5,6}$,
Emma Esparza-Borges$^{\orcidlink{0000-0002-2341-3233},7,8}$,
\newauthor
Chloe E. Fisher$^{{\orcidlink{0000-0003-0652-2902}},9}$,
Mercedes L\'opez-Morales$^{{\orcidlink{0000-0003-3204-8183}},10}$,
N. J. Mayne$^{{\orcidlink{0000-0001-6707-4563}},11}$,
Mason McCormack$^{{\orcidlink{0000-0002-1463-9847}},12}$,
\newauthor
Annabella Meech$^{{\orcidlink{0000-0002-7500-7173}},10,9}$,
Vatsal Panwar$^{\orcidlink{0000-0002-2513-4465},5,6}$,
Diana Powell$^{\orcidlink{0000-0002-4250-0957},12}$,  
Denis E. Sergeev$^{{\orcidlink{0000-0001-8832-5288}},11}$,
Jake Taylor$^{{\orcidlink{0000-0003-4844-9838}},9}$,
\newauthor
Shang-Min Tsai$^{\orcidlink{0000-0002-8163-4608},13}$,
Daniel Valentine$^{{\orcidlink{0000-0002-2643-6836}},4}$, 
Hannah R. Wakeford$^{{\orcidlink{0000-0003-4328-3867}},4}$,
Peter J. Wheatley$^{{\orcidlink{0000-0003-1452-2240}},5,6}$,
\newauthor
and Maria Zamyatina$^{{\orcidlink{0000-0002-9705-0535}},11}$
\\
$^{1}$Department of Physics, Imperial College London, Prince Consort Road, SW7 2AZ, London, UK \\
$^{2}$Max Planck Institute for Astronomy (MPIA), K\"{o}nigstuhl 17, 69117 Heidelberg, Germany \\
$^{3}$School of Physics and Astronomy, University of Leeds, Leeds LS2 9JT, UK\\
$^{4}$School of Physics, University of Bristol, HH Wills Physics Laboratory, Tyndall Avenue, Bristol BS8 1TL, UK \\
$^{5}$Centre for Exoplanets and Habitability, University of Warwick, Gibbet Hill Road, Coventry CV4 7AL, UK\\
$^{6}$Department of Physics, University of Warwick, Gibbet Hill Road, Coventry CV4 7AL, UK\\
$^{7}$Instituto de Astrof\'isica de Canarias, San Crist\'obal de La Laguna, Tenerife E-38200, Spain \\
$^{8}$Departamento de Astrof\'isica, Universidad de La Laguna, San Cristóbal de La Laguna, Tenerife E-38200, Spain\\
$^{9}$Department of Physics, University of Oxford, Keble Road, Oxford, OX1 3RH, UK\\
$^{10}$Center for Astrophysics ${\rm \mid}$ Harvard {\rm \&} Smithsonian, 60 Garden St, Cambridge, MA 02138, USA\\
$^{11}$Department of Physics and Astronomy, Faculty of Environment, Science and Economy, University of Exeter, Exeter EX4 4QL, UK \\
$^{12}$Department of Astronomy and Astrophysics, University of Chicago, IL, 60657, USA\\
$^{13}$Department of Earth and Planetary Sciences, University of California, Riverside, CA, USA \\
}

\date{Accepted 08/10/2024. Received 04/10/2024; in original form 04/07/2024}

\pubyear{\the\year{}}

\begin{document}
\label{firstpage}
\pagerange{\pageref{firstpage}--\pageref{lastpage}}
\maketitle

\begin{abstract}

A primary objective of exoplanet atmosphere characterisation is to learn about planet formation and evolution, however, this is challenged by degeneracies. To determine whether differences in atmospheric composition can be reliably traced to differences in evolution, we are undertaking a transmission spectroscopy survey with JWST to compare the compositions of a sample of hot Jupiters that have different orbital alignments around F stars above the Kraft break. Under the assumption that aligned planets migrate through the inner disc, while misaligned planets migrate after disc dispersal, the act of migrating through the inner disc should cause a measurable difference in the C/O between aligned and misaligned planets. We expect the amplitude and sign of this difference to depend on the amount of planetesimal accretion and whether silicates accreted from the inner disc release their oxygen. Here, we identify all known exoplanets that are suitable for testing this hypothesis, describe our JWST survey, and use noise simulations and atmospheric retrievals to estimate our survey's sensitivity. With the selected sample of four aligned and four misaligned hot Jupiters, we will be sensitive to the predicted differences in C/O between aligned and misaligned hot Jupiters for a wide range of model scenarios.

\end{abstract}

\begin{keywords}
Instrumentation -- Data methods -- JWST -- Exoplanets -- Planet formation -- Planet migration
\end{keywords}



\section{Introduction}

It has long been proposed that measuring an exoplanet’s atmospheric composition (specifically its carbon-to-oxygen ratio, C/O) can reveal information regarding where a planet formed with respect to different ice lines \citep[e.g.,][]{Oberg2011,Madhusudhan2014,Booth2017,Schneider2021}. The basic principle that the C/O of a planet's atmosphere is dependent on where it accreted its atmosphere relative to different C- and O-bearing molecular ice lines is robust, and there is little doubt atmospheric composition will lead to insights into planet formation and evolution. However, there are many uncertainties when relating an individual planet’s composition to its formation location, which, combined with our lack of sensitivity to carbon-bearing molecules in the pre-JWST era, have prevented a detailed investigation of how atmospheric composition depends on planet formation and evolution.

The challenges to our understanding of the link between formation and composition include the uncertain and evolving locations of ice lines within discs \citep[e.g.,][]{Morbidelli2016,Panic2017,Owen2020}, the observed diversity of protoplanetary discs \citep[e.g.,][]{Law2021}, how much solid versus gaseous material is accreted during planet formation \citep[e.g.,][]{Espinoza2017}, and the drift of solids relative to the gas in the disc \citep[e.g.,][]{Booth2017}. Furthermore, transit spectroscopy observations of exoplanets' atmospheres probe the atmospheric composition at the planetary limb which might hold inhomogeneities caused by, e.g., local atmospheric mixing \citep[e.g.,][]{Zamyatina24_quenchingdriven} or cloud formation \citep[e.g.,][]{Helling2016}, therefore not necessarily reflecting the bulk planet's atmospheric composition \citep{Muller2024}.

\begin{figure*}
    \centering
    \includegraphics[width=2\columnwidth]{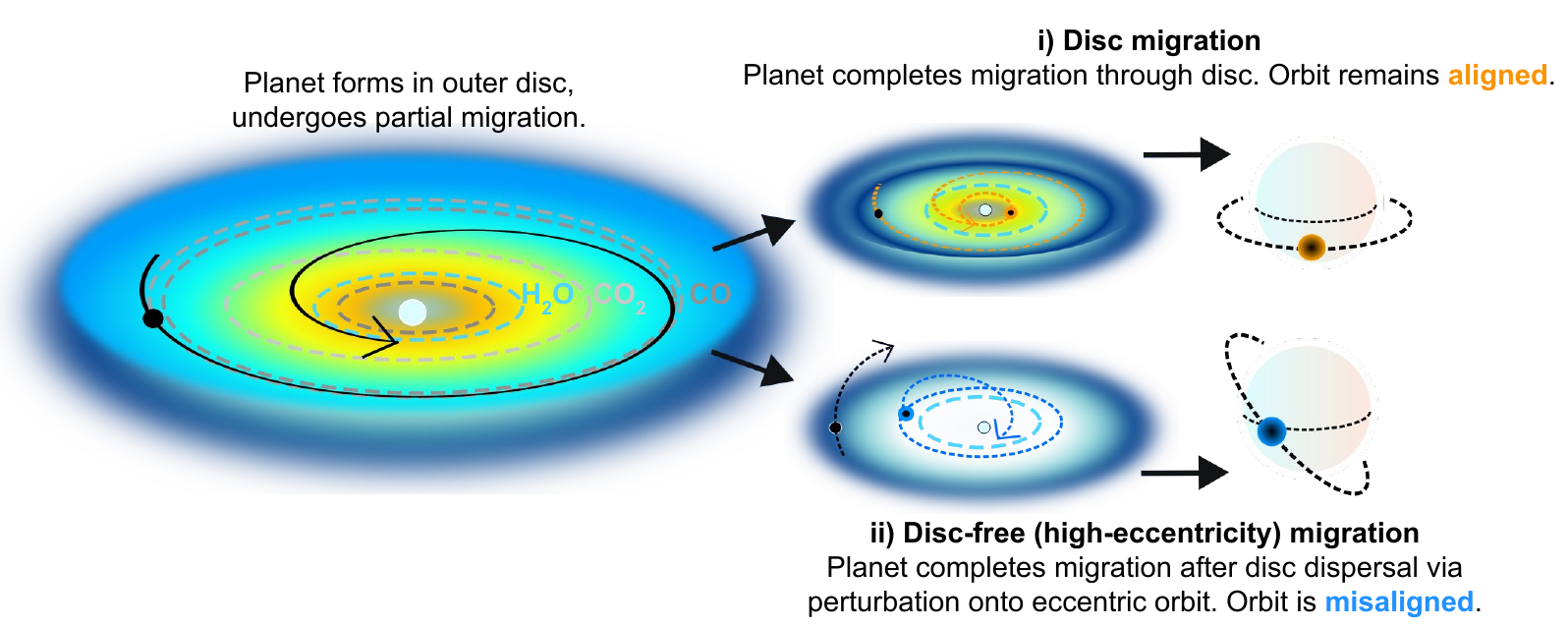}
    \caption{Schematic demonstrating the hypothesis test. The left hand diagram shows a planet that is born far out in the disc before undergoing partial migration through ice lines (CO and CO$_2$ in this example). Subsequently, there are two possible pathways this planet might follow to become a hot Jupiter: i) migration through the inner disc whereby the planet's orbit remains aligned (top right). In this case, the planet migrates through the H$_2$O ice line (light blue dashed circle) and accretes O-rich solids from the inner disc. ii) disc-free (high-eccentricity) migration, whereby the planet is perturbed onto an eccentric and misaligned orbit before undergoing tidal circularisation. This migration occurs after disc dispersal and so the planet does not accrete the O-rich solids from the inner disc.}
    \label{fig:migration_schematic}
\end{figure*}

\citetalias{PenzlinBooth2024} \citeyear{PenzlinBooth2024} used simulations to demonstrate that the unconstrained values of key disc and planet formation parameters such as dust-to-gas mass, disc temperature, and the relative drift of dust to gas (Stokes-to-alpha number), create a degeneracy between C/O and [O/H] (which we refer to as metallicity, Z, throughout this paper) over a wide dynamic range. These uncertainties are hard to constrain robustly through independent observations. Therefore, we propose the best way to determine whether differences in formation history lead to a measurable difference in atmospheric C/O and metallicity is by comparing populations of planets for which we have independent evidence that they underwent different evolutionary pathways. Specifically, as \citetalias{PenzlinBooth2024} \citeyear{PenzlinBooth2024} show, planets that have undergone disc-free (high eccentricity) migration should have different C/O and metallicity to planets that have undergone disc migration. The idea behind this is that disc-migrated planets will accrete solids from the inner disc during their migration while disc-free, high-eccentricity migrated planets will not since they complete their migration after disc dispersal (Figure \ref{fig:migration_schematic}).

With the advent of JWST's revolutionary precision and wavelength coverage of carbon-bearing molecules \citep[e.g.,][]{JWST2023,Alderson2023,Rustamkulov2023}, we are able to test these predictions for the first time against a well-designed target sample. To this end, we are undertaking a survey with JWST to compare the C/O and metallicity of four disc migrated hot Jupiters with four high eccentricity migrated hot Jupiters. We focus specifically on hot Jupiters, and not smaller planets, since their giant masses likely necessitate formation beyond ice lines and hence subsequent migration \citep[e.g.,][]{Lin1996,Rafikov2006,Dawson2018}. Furthermore, their massive H/He envelopes retain the primordial composition, without being changed by atmospheric loss \citep[e.g.,][]{Owen2018}. We will combine the transmission spectra of five planets from our new observational programme (GO 3838, 49.2 hours, PIs: Kirk \& Ahrer) with spectra of three planets from other programmes (GTO 1274, PI: Lunine; GTO 1353, PI: Lewis; GO 3154, PI: Ahrer). Our programme is called BOWIE-ALIGN, with BOWIE corresponding to the core institutions of our collaboration (Bristol, Oxford, Warwick, Imperial, Exeter) and ALIGN standing for A spectral Light Investigation into hot gas Giant origiNs. 

We distinguish disc migrated from high eccentricity migrated hot Jupiters via their sky-projected orbital alignments (`obliquities') around F stars where tidal realignment is thought to be inefficient \citep{Albrecht2012}. Disc migration is expected to lead to a slowly shrinking planetary orbit and the accretion of gas, dust and planetesimals in the migrating planet's path \citep[e.g.,][]{Dawson2018}. This results in little change in the eccentricity and inclination of a planet's orbital plane, which remains \textit{aligned} with the stellar spin axis (Figure \ref{fig:migration_schematic}). High eccentricity migration likely occurs after disc dispersal. Under this mechanism, it is thought that an initially cold Jupiter is perturbed into an eccentric orbit via interactions with a third body \citep[e.g.,][]{Rasio1996,Wu2003}, which drive up the eccentricity and inclination of the planet \citep[e.g.,][]{Kozai1962,Lidov1962,2016Munoz}. This method of migration is believed to result in \textit{misalignments} between a planet's orbital plane and the stellar spin axis. Therefore, by comparing aligned and misaligned hot Jupiters we can test the predicted impacts of migration method on atmospheric C/O and metallicity. 

The key with our survey is that since formation models are unable to a priori predict the specific values of C/O and metallicity for individual scenarios, they robustly predict a difference. As described by \citetalias{PenzlinBooth2024} \citeyear{PenzlinBooth2024}, the sign of this difference is even uncertain owing to uncertainties in formation models. Thus, by comparing one sample to another we can test this difference, along with narrowing down the range of uncertain disc parameters currently rendering the models unpredictive. Hence, rather than is the common expectation of measuring an atmospheric composition and comparing it to formation models to determine how the planet formed, we are proposing an opposite approach. Namely, testing the idea that different formation scenarios predict different atmospheric compositions, then using our measured compositions to constrain the formation and evolution models. 

Our paper is set out as follows: in Section \ref{sec:target_selection} we describe our target selection, in Section \ref{sec:observing_strategy} we describe the strategy behind our JWST observations, in Section \ref{sec:model_predictions} we explain our simulated transmission spectra and retrievals that are informed by coupled disc chemistry and planet formation models, in Section \ref{sec:data_reduction} we discuss our data reduction and analysis plans, in Section \ref{sec:existing_results} we discuss existing literature results for two planets in the sample, and in Section \ref{sec:ancillary_science} we describe additional science that will be enabled by our homogeneously observed sample. We summarise in Section \ref{sec:summary}.

\section{Target selection}
\label{sec:target_selection}

To test whether atmospheric composition depends on migration history, we needed to construct a sample of planets with large predicted atmospheric signals and precisely measured obliquities. To do this, we drew from the TEPCAT catalogue\footnote{https://www.astro.keele.ac.uk/jkt/tepcat/obliquity.html, accessed 23rd May 2024} \citep{Southworth2011} of 180 exoplanets with measured sky-projected obliquities, $\lambda$, primarily measured via the Rossiter-McLaughlin effect \citep{Rossiter1924,McLaughlin1924}. We define misaligned exoplanets as those with sky-projected obliquities \obliquity\ $> 45$\,\degree\ and aligned planets as those with \obliquity\ $< 30$\,\degree\ and only considered those planets with precisely measured obliquities ($\sigma$(\obliquity) $< 15$\,\degree). While we based our sample selection on sky-projected obliquities, it is possible that planets with low sky-projected obliquities could, in fact, be misaligned in 3D space once the stellar inclination is resolved. However, of the 20 planets with measured sky-projected \emph{and} 3D obliquities from the TEPCAT catalogue \citep{Southworth2011}, only one (5\,\% of the total) is aligned in 2D, sky-projected space and is misaligned in 3D space. This is encouraging for our target selection, which is based on sky-projected obliquities. Nevertheless, our sky-projected aligned sample could include planets that are misaligned in 3D space.

Importantly, we only considered planets that orbit stars above the Kraft break, defined as a sharp decrease in rotational velocity of stars due to tidal dissipation in their convective envelopes \citep{Kraft1967}. Stars below the Kraft break have deeper convective envelopes, so tidal-realignment is more efficient \citep[e.g.,][]{Albrecht2012,Spalding2022}. In other words, planets with low obliquities orbiting stars below the Kraft break may have tidally realigned after high-eccentricity migration, and therefore, it is believed that their obliquities are independent of their migration method. By choosing hot stars with radiative envelopes, we can be confident that hot Jupiters have retained their primordial obliquities. In practice, this restricts us to only F and A stars. We chose to consider only F stars (stellar effective temperature, $\mathrm{T_{eff}} \leq 7400$\,K) to limit potential atmospheric differences arising from differing stellar XUV histories and photodissociation rates.

As explained in the introduction, we chose to focus only on hot Jupiters (planet mass, \Mp\ > 0.3\,\Mjup, planet radius, \Rp\ > 0.5\,\Rjup, 1000 < \Teq\ < 2050\,K, where \Teq\ is planet equilibrium temperature) since they are likely to have formed beyond ice lines before migrating to their current locations \citep[e.g.,][]{Lin1996,Rafikov2006}. The upper bound on the equilibrium temperature was chosen to make sure that H$_2$O would not be thermally dissociated, which could complicate our inferences of the O abundance and hence the determination of C/O \citep[e.g.,][]{Kreidberg2018,Lothringer2018}. The value of 2050\,K was chosen because HST/WFC3 observations of WASP-19b (\Teq\ $= 2077 \pm 34$\,K, \citealt{Bernabo2024}) revealed significant water absorption without dissociation \citep{Huitson2013}. The lower temperature bound was to avoid including warm Jupiters that may undergo a different migration pathway to hot Jupiters \citep[e.g.,][]{Petrovich2016,Jackson2021}. 

While hot Jupiters have large amplitude atmospheric signals, the signal to noise will also depend on the brightness of the host star. For this reason, we made a final cut based on the Transmission Spectroscopy Metric \citep[TSM,][]{Kempton2018}, which accounts for the stellar magnitude. We chose to only consider planets with a K band TSM $>100$. 

After these cuts, we were left with a potential 16 hot Jupiters, comprising seven aligned and nine misaligned planets. As we show in Section \ref{sec:model_predictions}, adding more low noise planets does not significantly improve programme sensitivity. The list of potential targets is given in Table \ref{tab:sample}. From this list, we sub-selected the targets with the highest TSM signals. We did not select WASP-79b due to previous observations of stellar contamination in its transmission spectrum \citep{Rathcke2021} and did not include HAT-P-41b among our aligned sample owing to its relatively higher (more misaligned) obliquity ($\lambda = -22.1^{+0.8}_{-6.0}$\degree, \citealt{Johnson2017}). 

\begin{figure}
    \centering
    \includegraphics[scale=0.52]{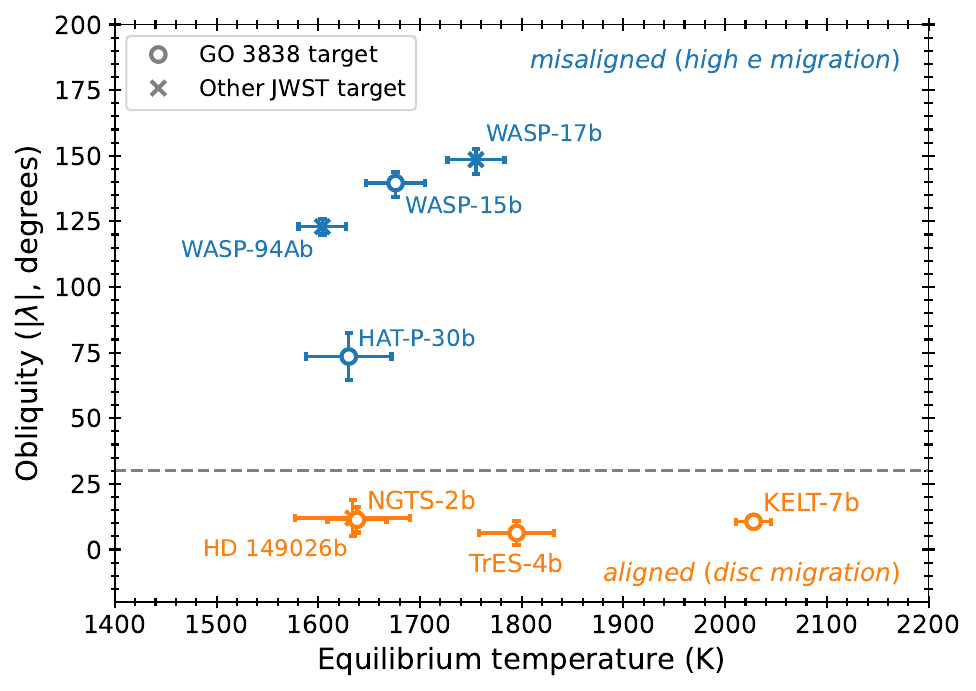}
    \includegraphics[scale=0.52]{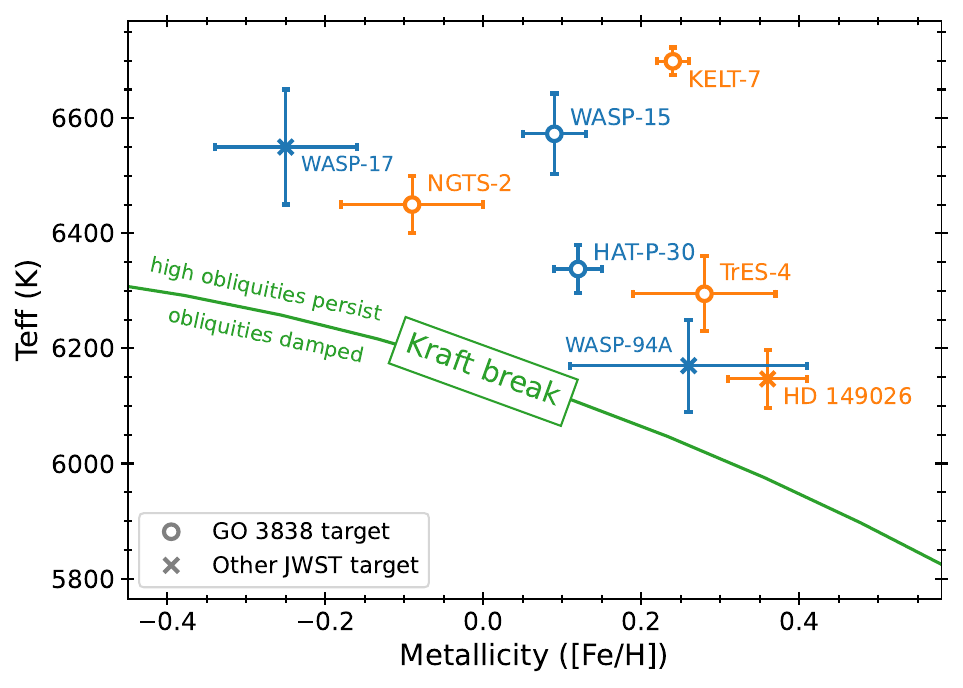}
    \caption{Top-panel: the obliquities and equilibrium temperatures of our sample. The aligned targets are shown in orange and misaligned targets are in blue. Open circles correspond to targets in JWST programme GO 3838, crosses to targets from other JWST programmes. Bottom-panel: the effective temperatures and metallicities of the host stars, plotted with respect to the metallicity-dependent Kraft break which is taken from \protect\cite{Spalding2022}.}
    \label{fig:sample_params}
\end{figure}

Our final list of targets is: TrES-4b, KELT-7b, HD\,149026b, NGTS-2b (all aligned), WASP-94Ab, WASP-17b, HAT-P-30b and WASP-15b (all misaligned). The first panel of Figure \ref{fig:sample_params} shows the planetary equilibrium temperatures and sky-projected obliquities of our sample, while the second panel shows the stellar effective temperatures and metallicities ([Fe/H]) along with the metallicity-dependent Kraft break from \cite{Spalding2022}. These planets span equilibrium temperatures of 1604--2028\,K, masses of 0.368--1.280\,\Mjup\ and radii of 0.813--1.932\,\Rjup. While we are primarily interested in how their atmospheric composition depends on obliquity and hence migration, this uniform sample of hot Jupiters orbiting F stars will allow for new insights into hot Jupiter atmospheres in general and enable several ancillary science cases, some of which we address in Section \ref{sec:ancillary_science}. 

Of these planets, we are acquiring new JWST observations of TrES-4b, KELT-7b, NGTS-2b, HAT-P-30b and WASP-15b through Programme GO 3838 (PIs: Kirk \& Ahrer) with the other targets already observed or due to be observed by other approved programmes (WASP-17b, GTO 1353, PI: Lewis; HD\,149026b, GTO 1274, PI: Lunine, \citealt{Bean2023}; WASP-94Ab, GO 3154, PI: Ahrer). All of these planets have eccentricities consistent with zero. With this combination of archival and new data, we will obtain (i) a four planet sample of aligned planets that likely migrated through the disc and (ii) a four planet sample of misaligned planets that ended their formation at orbits near and beyond the water ice line, and reach their final orbits after disc dispersal.

We note that our choice to divide aligned and misaligned planets at \obliquity\ $= 30$\,\degree\ was motivated by \cite{Spalding2022}'s definition. If we instead chose a more conservative definition of aligned planets, those with \obliquity\ $< 20$\,\degree, then HAT-P-41b (\obliquity\ $=22.10$\,\degree), XO-6b (\obliquity\ $=20.70$\,\degree) and WASP-3b (\obliquity\ $=20.0$\,\degree) would be classified as misaligned planets. Since these are not included in our observational sample, our observations are not sensitive to this definition of alignment.

A caveat to our experiment is the assumption that aligned planets are the result of disc migration and misaligned planets are the result of disc-free migration. However, there are other migration scenarios that could result in these obliquities. For example, coplanar high-eccentricity migration \citep{Petrovich2015} can occur in planetary systems with two or more gas giants. In this case, both planets share a low mutual inclination that remains aligned with the stellar spin axis. The inner planet can become an aligned, circularised hot Jupiter while the outer planet is typically 1--3 times more massive and remains on a moderately eccentric orbit ($e \sim 0.2-0.5$). \cite{Zink2023} used the California Legacy Survey to show that massive companions to hot Jupiters are ubiquitous at a population level. Furthermore, the eccentricity and mass distributions of these outer companions are compatible with the coplanar high-eccentricity migration scenario. While none of our aligned planets have a detected outer companion, outer companions could exist below the detection threshold. Separately, an inclined protoplanetary disc could result in misaligned hot Jupiters that underwent disc migration \citep[e.g.,][]{Batygin2012,Spalding2015}. We discuss the possible implications of these alternative scenarios in Section \ref{sec:summary}.

\begin{table*}
\caption{The list of 16 hot Jupiters that passed our selection criteria and are therefore suitable to address our science question. The eight planets in bold are those in our JWST survey. Each list is ordered by the K-band TSM \protect\citep{Kempton2018}. See the text for a description of these parameters. The parameter values are taken from TEPCAT \protect\citep{Southworth2011}. The following references for each planet are in the order discovery paper, most recent detailed study (as determined by TEPCAT), and obliquity: TrES-4b \protect\citep{Mandushev2007,Sozzetti2015,Narita2010}; KELT-7b \protect\citep{Bieryla2015,Tabernero2022}; HAT-P-41b \protect\citep{Hartman2012,Johnson2017}; HD\,149026b \protect\citep{Sato2005,Carter2009,Albrecht2012}; NGTS-2b \protect\citep{Raynard2018,Anderson2018}; XO-6b \protect\citep{Crouzet2017,Ridden-Harper2020}; WASP-3b \protect\citep{Pollacco2008,Maciejewski2013,Oshagh2013}; WASP-94Ab \citep{Neveu-VanMalle2014,Ahrer2024}; WASP-17b \protect\citep{Anderson2010,Triaud2010,Southworth2012}; HAT-P-30b \protect\citep{Johnson2011,Blazek2022,Cegla2023}, WASP-79b \protect\citep{Smalley2012,Brown2017}; WASP-15b \protect\citep{West2009,Southworth2013,Triaud2010}; WASP-7b \protect\citep{Hellier2009,Southworth2012b,Albrecht2012}; WASP-109b \protect\citep{Anderson2014,Addison2018}; HAT-P-6b \protect\citep{Noyes2008,Southworth2012b,Albrecht2012}; WASP-180b \protect\citep{Temple2019}. Note: HAT-P-41b, XO-6b and WASP-3b would be classified as misaligned planets if the criteria for misalignment was defined as \obliquity\ $> 20$\,\degree. They are not included in our observational sample.}
\label{tab:sample}
\begin{tabular}{lcccccccc}
\hline 
Planet &  $\mathrm{T_{eff}}$ (K) & [Fe/H] &  $\lambda$ ($^\circ$) & $\mathrm{M_P}$ ($\mathrm{M_{Jup}}$) &  $\mathrm{R_P}$ ($\mathrm{R_{Jup}}$) &  $\mathrm{T_{eq}}$ (K) &  K magnitude &  TSM$_K$ \\ \hline 
\multicolumn{9}{l}{\color{orange}\textit{Aligned sample}} \\
\textbf{TrES-4b}$^{a}$    &    $ 6295 \pm 65 $ &  $ 0.280 \pm 0.090 $ &   $ 6.30 \pm 4.70 $ & $ 0.494 \pm 0.035 $ & $ 1.838 \pm 0.086 $ & $ 1795 \pm 37 $ & 10.3 &  302   \\
\textbf{KELT-7b}$^{a}$   &   $ 6699 \pm 24 $ & $ 0.240 \pm 0.020 $ &  $ -10.55 \pm 0.27 $ & $ 1.280 \pm 0.170 $ &  $ 1.496 \pm 0.035 $ & $ 2028 \pm 17 $ &  7.5 &  287  \\
HAT-P-41b  &   $ 6390 \pm 100 $ &  $ 0.210 \pm 0.100 $ &  $ -22.10^{+0.80}_{-6.00} $ & $ 0.800 \pm 0.1020 $ & $ 1.685 \pm 0.064 $ & $ 1941 \pm 38 $ &   9.7 &  237  \\
\textbf{HD 149026b}$^{b}$ &   $ 6147 \pm 50 $ & $ 0.360 \pm 0.050 $ &  $ 12.00 \pm 7.00 $ & $ 0.368 \pm 0.014 $ &  $ 0.813 \pm 0.026 $ & $ 1634 \pm 57 $ &  6.8 &  221  \\
\textbf{NGTS-2b}$^{a}$   &   $ 6450 \pm 50 $ & $ -0.090 \pm 0.090 $ &  $ -11.30 \pm 4.80 $ &  $ 0.670 \pm 0.089 $ &  $ 1.536 \pm 0.062 $ & $ 1638 \pm 29 $ &   9.8 &  189  \\
XO-6b      &   $ 6720 \pm 100 $ & $ -0.070 \pm 0.100 $ &  $ -20.70 \pm 2.30 $ & $ 2.010 \pm 0.710 $ & $ 2.080 \pm 0.180 $ &  $ 1641 \pm 24 $ &   9.3 &  142  \\
WASP-3b  &   $ 6340 \pm 90 $ & $ 0.161 \pm 0.063 $ &   $ 20.0 \pm 3.3 $ & $ 1.770 \pm 0.100 $ &  $ 1.346 \pm 0.063 $ & $ 2020 \pm 35 $ &   9.4 &  113  \\ \hline
\multicolumn{9}{l}{\color{blue}\textit{Misaligned sample}} \\ 
\textbf{WASP-94Ab}$^{c}$ & $ 6170 \pm 80$ & $ 0.26 \pm 0.15 $ & $ 123 \pm 3 $ & $ 0.452 \pm 0.034 $ & $ 1.720 \pm 0.055 $ & $ 1604 \pm 24 $ & 8.9 & 590 \\
\textbf{WASP-17b}$^{d}$ &    $ 6550 \pm 100$ & $-0.25 \pm 0.09$ &  $-148.5^{+4.2}_{-5.4}$ &  $0.477 \pm    0.033$ &  $ 1.932 \pm    0.053 $ &  $ 1755 \pm     28 $ &  10.2 &  488  \\
\textbf{HAT-P-30b}$^{a}$ &    $ 6338 \pm 42$ &  $0.12 \pm 0.03$ &   $ 70.5^{+2.9}_{-2.8} $ &  $0.711 \pm    0.028$ &  $1.417 \pm    0.033$ &  $1630 \pm     42$ &   9.2 &  333  \\
WASP-79b &    $ 6600 \pm 100$ &  $0.03 \pm 0.10$ &   $-95.2^{+0.9}_{-1.0}$ &  $0.860 \pm    0.080$ &  $1.530 \pm    0.040$ &  $1716 \pm     25$ &   9.1 &  246  \\
\textbf{WASP-15b}$^{a}$ &   $ 6573 \pm 70$ &  $0.09 \pm 0.04$ &  $-139.6^{+4.3}_{-5.2}$ &  $0.592 \pm    0.019$ &  $1.408 \pm    0.046$ &  $1676 \pm     29$ &   9.7 &  201  \\
WASP-7b &    $ 6520 \pm 70$ &  $0.00 \pm 0.10$ &    $86.0 \pm 6.0$ &  $0.980  \pm 0.130$ & $1.374 \pm    0.094$ &  $1530 \pm     45$ &   8.4 &  198  \\
WASP-109b &    $ 6520 \pm 140$ & $-0.22 \pm 0.08$ &    $99.0^{+10.0}_{-9.0}$ &  $0.910 \pm    0.130$ &  $1.443 \pm    0.053$ &  $1685 \pm     40$ &  10.2 &  143  \\
HAT-P-6b &    $ 6570 \pm 80$ & $-0.13 \pm 0.08$ &   $165.0 \pm 6.0$ &  $1.063 \pm 0.057$ & $1.395 \pm   0.081$ &  $1704 \pm     40$ &   9.3 &  132  \\
WASP-180b &    $ 6500 \pm 150$ &  $0.10 \pm 0.20$ &  $-162.0 \pm 5.0$ &  $0.900 \pm 0.100$ & $1.240 \pm 0.040$ &  $1540 \pm     40$ &   9.8 &  128  \\
\hline
\multicolumn{9}{l}{$^a$Observed by our new JWST programme, GO 3838 (PIs: Kirk \& Ahrer)} \\
\multicolumn{9}{l}{$^b$Observed in emission as part of JWST GTO 1274 (Lunine) and published in \protect\cite{Bean2023}} \\
\multicolumn{9}{l}{$^c$Observed as part of JWST GO 3154 (PI: Ahrer)} \\
\multicolumn{9}{l}{$^d$Observed as part of JWST GTO 1353 (PI: Lewis)} \\
\end{tabular}
\end{table*}

\section{Observing strategy}
\label{sec:observing_strategy}

\begin{figure}
\centering
\includegraphics[scale=0.47]{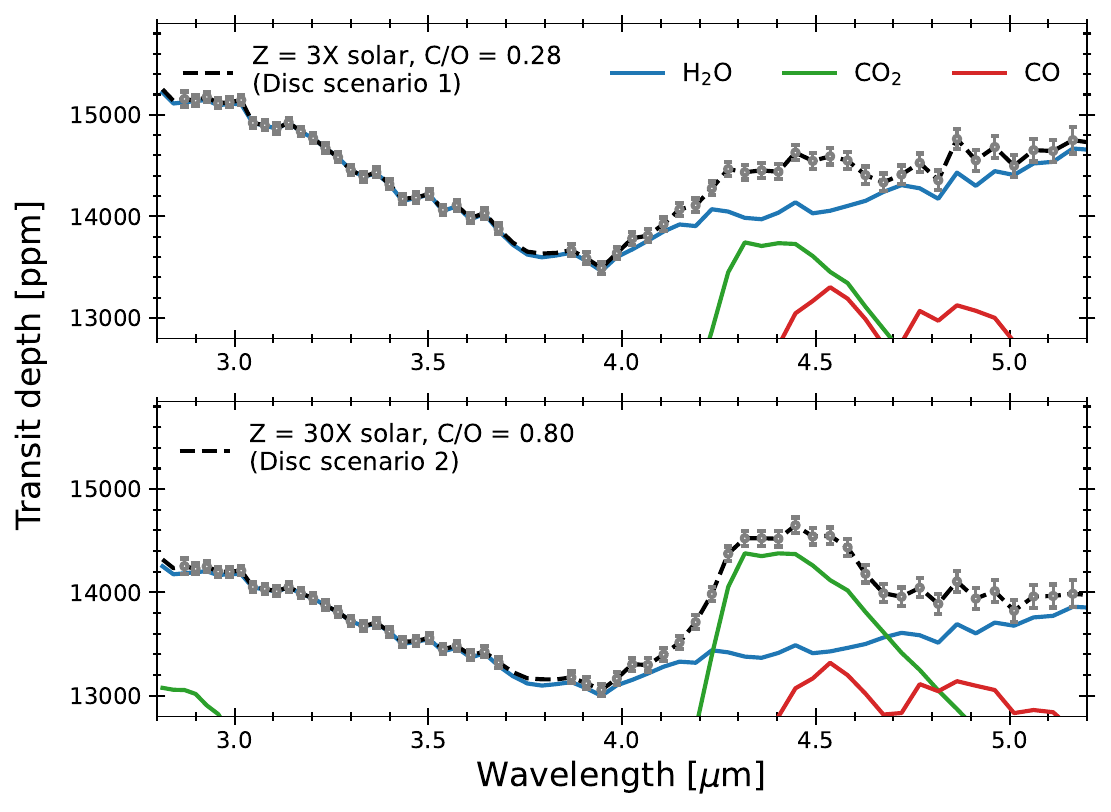}
\caption{Two example forward models used in this analysis, both created for TrES-4b (an aligned target). Top panel: a $3\times$ solar metallicity, $\mathrm{C/O} = 0.28$ forward model (black dashed line), corresponding to the fiducial disc model (disc scenario 1, Section \ref{sec:model_predictions} and Figure \ref{fig:anna_models}, left panel). The grey uncertainties correspond to the estimated JWST precision at a spectral resolution of $R=100$. The colour lines show the contribution of different species to the overall opacity, using line lists for H$_2$O from \protect\cite{Polyansky2018}, CO$_2$ from \protect\cite{Tashkun2011}, and CO from \protect\cite{Faure2013,Gordon2017}. Bottom panel: a $30\times$ solar metallicity, $\mathrm{C/O} = 0.80$ forward model (black line), corresponding to the silicate rainout disc model (disc scenario 2, Section \ref{sec:model_predictions} and Figure \ref{fig:anna_models}, right panel).}
\label{fig:opacities}
\end{figure}

We will observe our targets in transmission using the JWST NIRSpec G395H instrument mode, which covers several absorption bands from the primary O and C bearing species in hot Jupiters (H$_2$O, CO$_2$ and CO, Figure \ref{fig:opacities} and e.g., \citealt[][]{Alderson2023}). This approach will allow us to measure C/O and O/H and thus address whether these quantities differ between aligned and misaligned hot Jupiters. While NIRSpec/PRISM would cover the same molecular features in addition to covering bluer wavelengths (albeit at lower spectral resolution), it saturates more quickly than NIRSpec/G395H. The JWST Early Release Science observations of WASP-39b (K mag = 10.2) with NIRSpec/PRISM saturated at near-IR wavelengths \citep{Rustamkulov2023}, with the data synthesis analysis recommending the avoidance of partial saturation with PRISM \citep{Carter2024}. Since four of our five GO 3838 targets are brighter than WASP-39 (Table \ref{tab:sample}), we selected G395H to avoid saturation and to ensure a homogeneous data set across the sample. Similarly, the observations of WASP-17b and WASP-94Ab also make use of the G395H instrument. However, HD 149026b was observed with NIRCam in emission \citep{Bean2023}, and therefore, we will need to take this into account when we combine the full sample. We discuss the results of \cite{Bean2023} in Section \ref{sec:existing_results}.

We will observe single transits of our planets in Bright Object Time Series mode with the 2048 subarray and F290LP filter, with each observation consisting of $10^2$--$10^3$ integrations covering a continuous baseline of several hours. This setup will provide high-precision spectrophotometry covering the wavelength range 3--5\,\micron~at $R=2700$, allowing us to measure the planet's wavelength-dependent transit depth (its `transmission spectrum'). For all of our targets, we set the number of groups per integration to fill 80\,\% of the full well. Our observation duration per target was set to be equal to each planet's transit duration plus an additional four hours. These four hours comprise a minimum pre- and post-transit baseline of 1.5 hours to enable accurate and precise relative transit depth measurements, along with a one hour window for scheduling flexibility. For the non-GO 3838 targets, the out-of-transit/eclipse baselines were 3.5 hours for WASP-94Ab (GO 3154), 5.5 hours for WASP-17b (GTO 1353) and 5 hours for HD\,149026b (GTO 1274).

As all our GO 3838 science targets are too bright to use them for target acquisition (TA), we will utilise nearby (within splitting distance), fainter stars to conduct our target acquisition via the Wide Aperture Target Acquisition (WATA) mode. We have selected them using both the 2MASS \citep{Skrutskie2006} and GAIA DR3 \citep{Gaia,GaiaDR3} catalogues. The first one ensures that we have the necessary brightness for each star to achieve successful TA (SNR > 20 as verified with the JWST Exposure Time Calculator, ETC), while the latter ensures that the positions and proper motions of each star are accurate. 

\section{Predictions from disc models and 1D atmosphere models}
\label{sec:model_predictions}

In order to determine to what significance we could infer the predicted difference between C/O and metallicity for aligned and misaligned planets, we generated a grid of 1D isothermal chemical equilibrium atmosphere models using the exoplanet atmosphere radiative transfer forward and retrieval modelling code \texttt{PLATON} \citep{Zhang2019,Zhang2020}. We used the $R=10000$ line lists and \texttt{PLATON}'s default species which, at the wavelengths and temperatures we are considering (Figure \ref{fig:opacities}), are dominated by H$_2$O \citep{Polyansky2018}, CO$_2$ \citep{Tashkun2011}, and CO \citep{Faure2013,Gordon2017}\footnote{See Table 4 of \cite{Zhang2020} for the full list of species included in \texttt{PLATON}.}. 

\begin{figure*}
    \centering
    \includegraphics[scale=0.575]{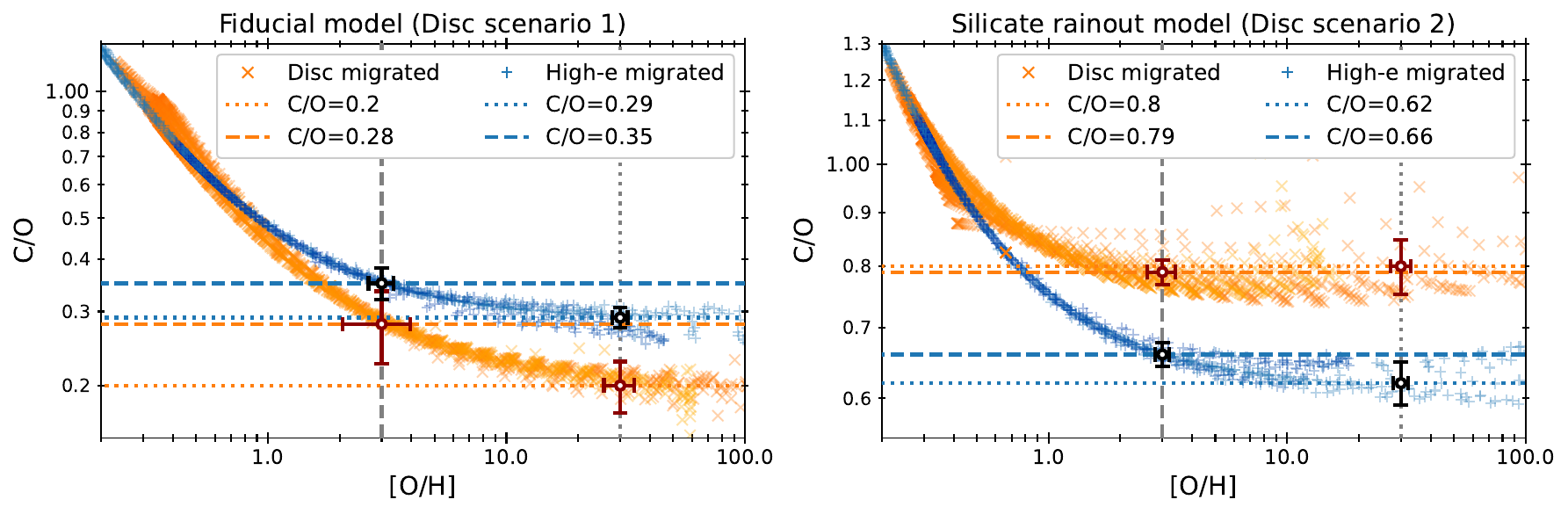}
    \caption{The C/O and [O/H] (= Z) results of two models from the suite of models run in \protect\citetalias{PenzlinBooth2024} \citeyear{PenzlinBooth2024}. Left panel: The C/O and [O/H] from the fiducial model (disc scenario 1), showing that aligned, disc-migrated planets (orange crosses) tend towards lower C/O than misaligned, high-eccentricity migrated planets (blue plusses) due to the accretion of O-rich ices in the inner disc. The dashed and dotted lines indicate the two values of metallicity and four values of C/O that were adopted in our simulation setup here. The white circles with uncertainties show the precision with which we'll be able to measure these values from an eight-planet sample. Specifically, the uncertainties are the result of co-adding four aligned planets (red error bars) and four misaligned planets (black error bars), assuming each planet has the same mean C/O and [O/H], as set by the dashed/dotted lines. Right panel: The C/O and [O/H] from the silicate rainout model (disc scenario 2), whereby silicates do not release their O into the atmospheres of exoplanets upon accretion. In this case, aligned, disc-migrated planets (orange points) tend towards higher C/O than misaligned, high-eccentricity migrated planets (blue points). Again, the dashed and dotted lines indicate the two values of metallicity and four values of C/O that were adopted in our simulation setup here.}
    \label{fig:anna_models}
\end{figure*}

Along with planetary parameters such as radius and temperature, the \texttt{PLATON} forward models are parameterised by metallicity (Z) and C/O. In its default configuration, \texttt{PLATON} assumes solar elemental abundances from \cite{Asplund2009} and a C/O ratio of 0.53. For our purposes, we set Z and C/O according to the findings of \citetalias{PenzlinBooth2024} \citeyear{PenzlinBooth2024}. Typically, planet formation and disc models have focused on Sun-like stars; however, the \citetalias{PenzlinBooth2024} \citeyear{PenzlinBooth2024} calculations were specifically designed for F stars like those in our survey, which have hotter discs. We considered the results of two end scenarios that are distinctly opposing, but representative: 

\begin{enumerate}
    \item[1)] the fiducial model whereby aligned planets trend towards \textit{lower} C/O than misaligned planets, owing to their accretion of O-rich ices from the inner disc and the fact that accreted silicates from the inner disc evaporate their O into the atmosphere (`disc scenario 1', Figure \ref{fig:anna_models}, left panel) and 
    \item[2)] the silicate rainout model whereby accreted silicates do not evaporate their oxygen into the atmospheres of planets. In this case, the atmospheric composition is dominated by gas, volatiles and carbon-rich refractories. This increases the C/O of disc migrated planets and thus causes aligned planets to have \textit{higher} C/O than misaligned planets (`disc scenario 2', Figure \ref{fig:anna_models}, right panel).
\end{enumerate}

With these two scenarios, we then selected two different metallicities ($3\times$ and $30\times$ solar, vertical lines in Figure \ref{fig:anna_models}) and the corresponding C/O ratios for both the aligned and misaligned planets (horizontal lines, Figure \ref{fig:anna_models}). We selected super-solar metallicities given the existing evidence from JWST for the prevalence of hot Jupiters with super-solar metallicity atmospheres \citep[e.g.,][]{Alderson2023,Ahrer2023,Bean2023,Feinstein2023,Rustamkulov2023,Xue2024}. In total, we considered eight combinations of Z and C/O: 2 disc scenarios (fiducial, silicate rainout) $\times$ 2 metallicities ($3\times$, $30\times$ solar) $\times$ 2 C/O (aligned, misaligned). We also made a second set of eight atmosphere models where we included a grey cloud deck at a pressure of 1\,mbar, which acts to obscure molecular absorption from pressures higher than 1\,mbar. We did this to consider the impact of muted molecular absorption on our ability to constrain Z and C/O. 

With the grid of 16 atmospheric forward models defined, we proceeded to generate simulated transmission spectra for all 16 planets in Table \ref{tab:sample} (for a total of 256 simulated transmission spectra). Figure \ref{fig:opacities} shows two example transmission spectra for TrES-4b, an aligned target. We added simulated error bars to our model spectra using the JWST noise software package \texttt{PandExo} \citep{Batalha2017}. For the seven planets in our sample that are scheduled for NIRSpec/G395H observations, we set the number of groups and integrations equal to the actual numbers used in the observations. For the five targets observed through GO 3838, and those targets not observed to date, the number of groups is set to get as close to, but not exceed, 80\,\% of the full well calculated from the JWST ETC, with the number of integrations set to cover the transit duration plus 4 hours, for the reasons given in Section \ref{sec:observing_strategy}. The example spectra in Figure \ref{fig:opacities} include measurement uncertainties generated by this method.

With the model spectra from \texttt{PLATON} and uncertainties from \texttt{PandExo}, we proceeded to use \texttt{PLATON}'s retrieval capabilities to determine the precision to which we expect to infer Z and C/O from our synthetic spectra. The free parameters in our retrievals were the planet's atmospheric metallicity ($\log Z$), atmospheric C/O, planet radius ($\mathrm{R_P}$) and the temperature of the isothermal atmosphere ($\mathrm{T_{iso}}$), plus a cloud-top pressure for the cloudy spectra. We placed flat, wide priors on all the retrieved parameters. Specifically, $\log Z$ was bounded between -1 and 3, C/O between 0.05 and 2.0, $\mathrm{R_P}$ between $0.9\times$ and $1.1\times$ the input value, and $\mathrm{T_{iso}}$ between 300 and 2500\,K. For models with clouds, the cloud-top pressure was bounded between $10^{-6}-1$\,bar. 

\begin{figure*}
    \centering
    \includegraphics[scale=0.8]{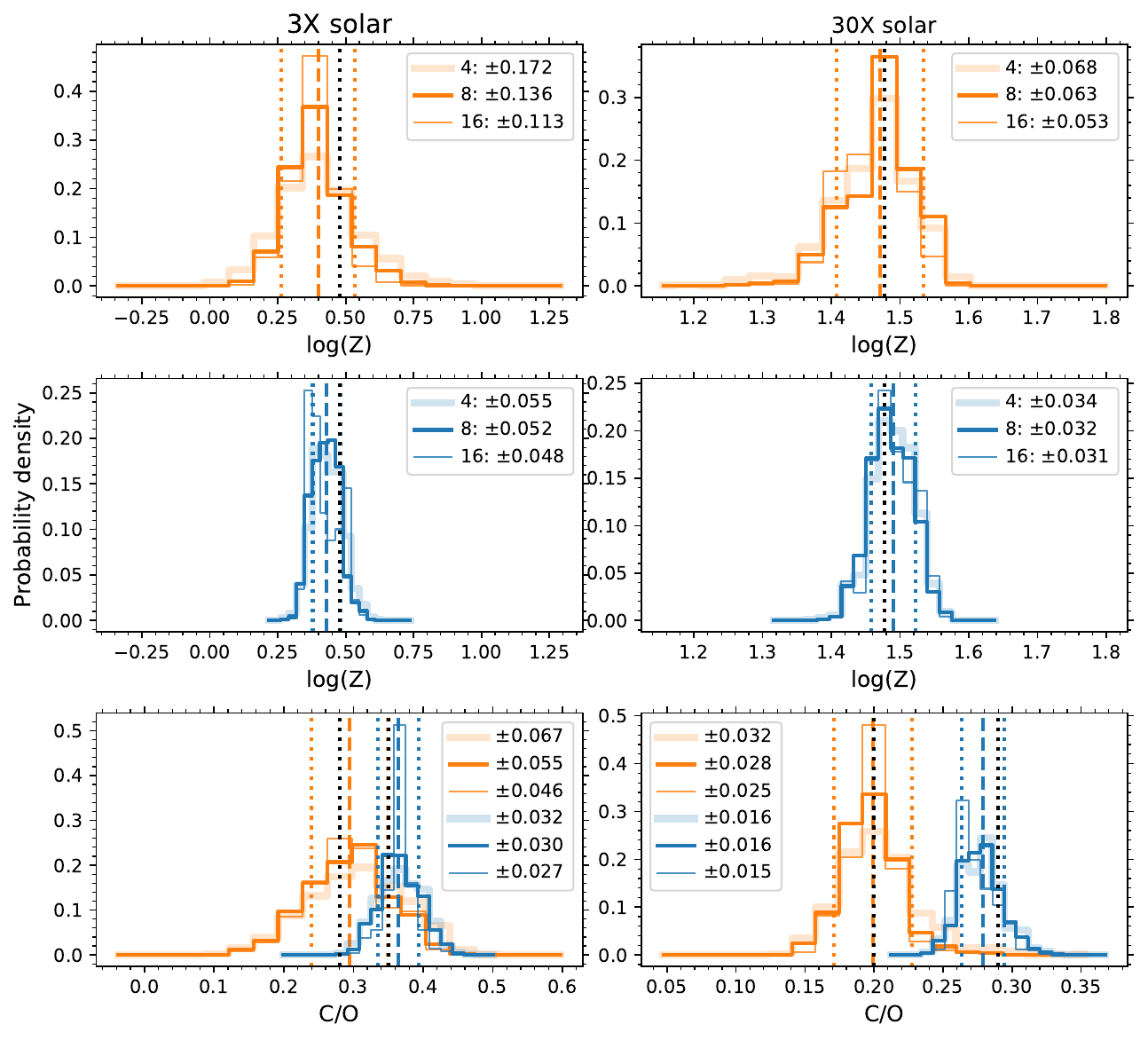}
    \caption{The retrieved $\log Z$ and C/O posteriors from the model atmospheres with $\log Z$ and C/O set by the fiducial disc model (scenario 1). The left column corresponds to the $3\times$ solar metallicity cases, and the right column to $30\times$ solar. Top row: the combined $\log Z$ posteriors from the aligned planets for a \emph{total} sample size of 4 (2 aligned, 2 misaligned), 8 (4 aligned, 4 misaligned) and 16 (7 aligned, 9 misaligned) planets. The legend gives the standard deviations of these distributions, which are equivalent to co-adding the uncertainties in $\log Z$ and C/O for 2 planets (thick, pale lines), 4 planets (thinner, dark lines) and 7 aligned / 9 misaligned planets (thinnest lines). The darkest line corresponds to our actual sample size of 8 planets. The vertical dashed and dotted orange lines indicate the mean and standard deviation of the 8 planet posterior. The vertical dotted black line indicates the input value. Second row: $\log Z$ posteriors for the misaligned planets. Bottom row: the C/O posteriors for both the aligned (orange) and misaligned planets (blue).}
    \label{fig:posteriors_fiducial}
\end{figure*}

\begin{figure*}
    \centering
    \includegraphics[scale=0.8]{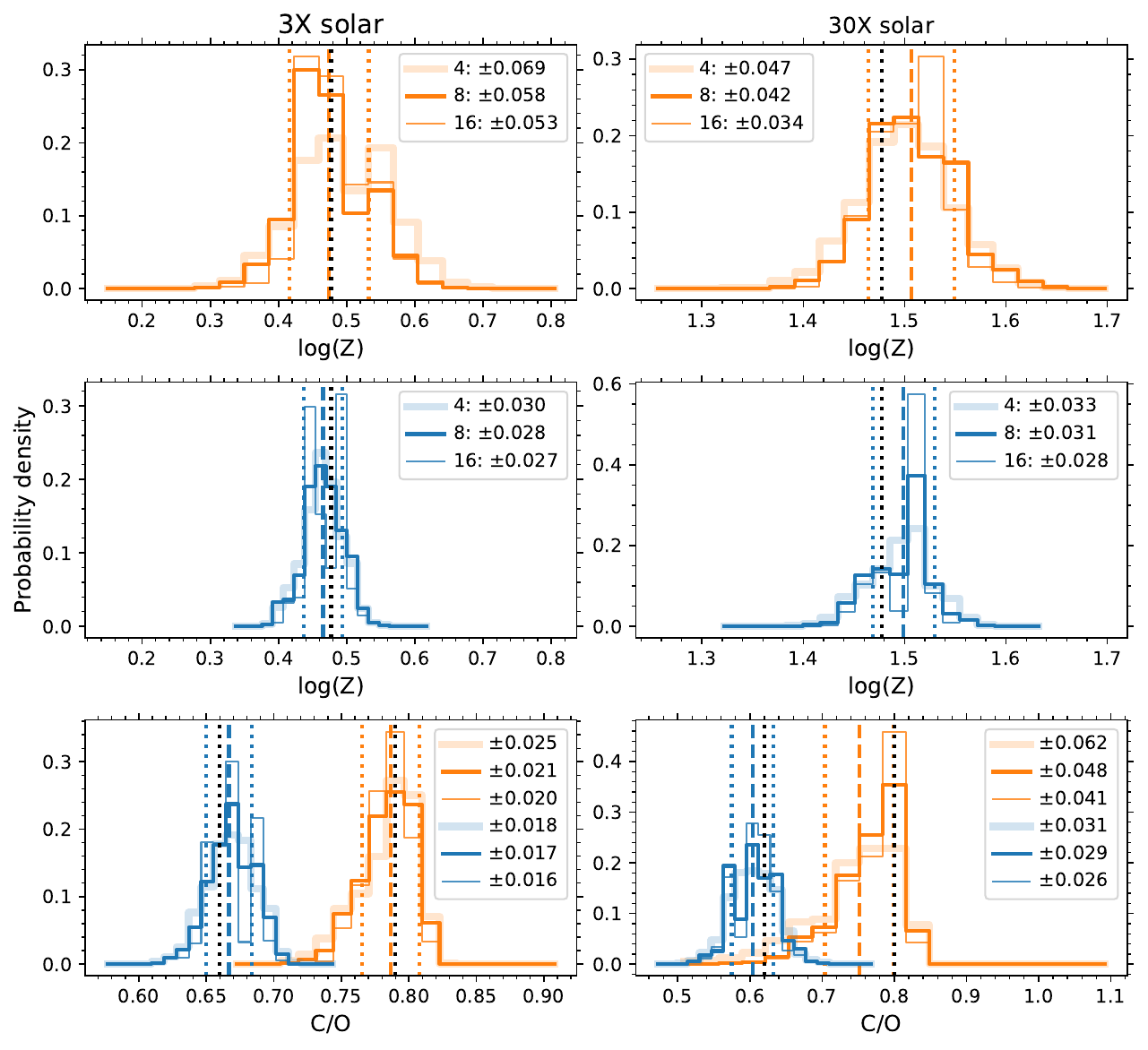}
    \caption{The retrieved $\log Z$ and C/O posteriors from the model atmospheres with $\log Z$ and C/O set by the silicate rainout disc model (scenario 2). See Figure \protect\ref{fig:posteriors_fiducial} for a description of the axes.}
    \label{fig:posteriors_noSi}
\end{figure*}

We sampled the parameter space using nested sampling, implemented via \texttt{dynesty} \citep{Speagle2020}, with 100 live points. To determine the benefit of increasing the sample size, we sequentially combined the posteriors of the best 4 (2 aligned, 2 misaligned), 8 (4 aligned, 4 misaligned) and 16 (7 aligned, 9 misaligned) planets, where the best is determined by the TSM ranking in Table \ref{tab:sample}. The posteriors for the fiducial model setup (disc scenario 1) are shown in Figure \ref{fig:posteriors_fiducial} with the silicate rainout model (disc scenario 2) shown in Figure \ref{fig:posteriors_noSi}. The posteriors resulting from the fits to the models with the grey cloud deck at 1\,mbar are shown in Appendix \ref{sec:apx_cloud}. 

As Figures \ref{fig:posteriors_fiducial} and \ref{fig:posteriors_noSi} show, there are diminishing returns as the sample size is increased. This is because to increase the total sample size, we are adding planets with lower signal-to-noise (as shown by their TSMs in Table \ref{tab:sample}). Also, we see that increasing the sample size leads to a more significant improvement in the $\log Z$ and C/O precisions for the aligned planets (orange histograms) as compared to the misaligned planets (blue histograms). This is due to the fact that the TSMs of the two best misaligned planets (WASP-94Ab and WASP-17b) are significantly higher than the remaining misaligned planets, while the two best aligned planets (TrES-4b and KELT-7b) are more similar to the rest of the aligned planets in terms of TSM (Table \ref{tab:sample}).

For our actual sample size of eight planets, we estimate a precision on C/O of $\pm 0.055$ (aligned planets) and $\pm 0.030$ (misaligned planets) for the $3\times$ solar metallicity, fiducial case (disc scenario 1). This would make the predicted difference between the aligned and misaligned planets of $\Delta(\mathrm{C/O}) = 0.07$ particularly challenging to measure ($< 1\sigma$). However, if the planets are more metal-rich, similar to other planets in the literature, then we would be sensitive to the difference between the C/O of aligned and misaligned planets in the $30\times$ solar metallicity fiducial setup. Specifically, we estimate a precision on C/O of $\pm 0.028$ (aligned planets) and $\pm 0.016$ (misaligned planets) in this case, which would allow us to measure the predicted difference between the C/O of the two populations of $\Delta(\mathrm{C/O}) = 0.09$ at $2.0\sigma$. 

For disc scenario 2, where silicates do not release their oxygen upon accretion, this leads to a larger predicted difference in C/O between the two populations that we'd have greater sensitivity to, with the aligned planets having higher C/O. In this scenario, we estimate a precision on C/O of $\pm 0.021$ (aligned planets) and $\pm 0.017$ (misaligned planets) for the $3\times$ solar metallicity atmospheres. These precisions would allow us to measure the predicted difference between the aligned and misaligned planets that do not undergo silicate evaporation of $\Delta(\mathrm{C/O}) = 0.13$ at $3.4\sigma$. For the $30\times$ solar metallicity atmospheres, we estimate a precision on C/O of $\pm 0.048$ (aligned planets) and $\pm 0.029$ (misaligned planets), allowing us to measure the predicted difference between the C/O of the two populations, $\Delta(\mathrm{C/O}) = 0.18$, at $2.3\sigma$.

As Figures \ref{fig:posteriors_fiducial_mbar_cloud} and \ref{fig:posteriors_noSi_mbar_cloud} show, the inclusion of a grey cloud deck widens the C/O posteriors by up to 0.037. Despite this, we would still be sensitive to the predicted difference in the C/O for $30\times$ metallicity atmospheres for both disc scenarios and the difference in C/O for the $3\times$ solar metallicity atmosphere for disc scenario 2.

Five of our targets will be observed with additional instruments in other approved JWST programmes. Specifically, WASP-17b will be observed with JWST/NIRISS, NIRSpec/G395H and MIRI/LRS covering wavelengths from $\sim 0.6-12$\,\micron\ in both transmission and emission (GTO 1353, PI: Lewis). HAT-P-30b, NGTS-2b, KELT-7b and WASP-94Ab will be observed with JWST/NIRISS in Cycle 3 (GO 5924, PI: Sing), providing transmission spectra from 0.6--2.8\,\micron. Combining these future data with our G395H observations will allow for tighter constraints than we predict here, although this will require careful analysis to avoid instrument-specific biases arising in atmospheric inferences \citep[e.g.,][]{Lueber2024}. 

\section{Data reduction and analysis strategy}
\label{sec:data_reduction}

To reduce and analyse our data, we will follow and learn from the approaches used in the JWST Early Release Science (ERS) Transiting Exoplanet Program \citep[][]{JWST2023,Alderson2023,Ahrer2023,Feinstein2023,Rustamkulov2023}. Specifically, we will reduce each of our data sets with more than one independent open-source pipeline, namely \texttt{Tiberius}\footnote{\url{https://github.com/JamesKirk11/Tiberius}} \citep{Kirk2017,Kirk2021}, \texttt{Eureka!}\footnote{\url{https://github.com/kevin218/Eureka}} \citep{Bell2022} and \texttt{ExoTiC-JEDI}\footnote{\url{https://github.com/Exo-TiC/ExoTiC-JEDI}} \citep{Alderson2022}. Each of these reduction pipelines has been used in the ERS programme and multiple subsequent JWST analysis papers. While they largely produce results consistent within uncertainties \citep[e.g.,][]{Alderson2023,Ahrer2023,Rustamkulov2023}, small differences between the resulting spectra can lead to differences in inferences regarding planetary atmospheres \citep[e.g.,][]{Constantinou2023,Kirk2024}. By comparing multiple reductions for every observation in the programme, we will determine how our conclusions depend on the choice of reduction pipeline and explore the origins of any differences. 

Data reduction (raw images to planet spectra) is only one half of the analysis needed to infer atmospheric C/O and Z. The second half involves going from planet spectra to atmospheric constraints. To perform this step, we will use at least two independent open-source forward modelling and retrieval codes to interpret each planet's spectrum, including \texttt{PLATON}\footnote{\url{https://github.com/ideasrule/platon}} \citep{Zhang2019,Zhang2020}, \texttt{petitRADTRANS}\footnote{\url{https://gitlab.com/mauricemolli/petitRADTRANS}} \citep{Molliere2019,Nasedkin2024}, \texttt{CHIMERA}\footnote{\url{https://github.com/mrline/CHIMERA}} \citep{Line2012,Line2013} and \texttt{VULCAN}\footnote{\url{https://github.com/exoclime/VULCAN}} \citep{Tsai2017,Tsai2021}. Each of these codes makes different modelling assumptions, including equilibrium vs.\  disequilibrium chemistry, isothermal vs.\ non-isothermal temperature profiles, and grey vs.\ non-grey clouds. As has been shown by the ERS programme (Welbanks et al., in prep.), different modelling assumptions can also influence the measured C/O and Z from a planet's spectrum which motivates our decision to use at least two codes per spectrum. 

Given the differences that can arise from both the data reduction and data interpretation steps, we will have at least one constant approach running through all of our planetary analyses. Specifically, a \texttt{Tiberius} reduction with the same extraction parameters and wavelength bins and a \texttt{petitRADTRANS} retrieval with the same gases and temperature--pressure profile parameterisation. This will avoid reduction-dependent biases, allowing us to fairly assess how the C/O differs between the aligned and misaligned planets in an unbiased way. 

Upon the conclusion of each planet's analysis, we will make multiple data products openly available on Zenodo. These data sets will include extraction input files, calibrated image files, extracted 2D stellar spectra, white and spectroscopic light curves, light curve models, transmission spectra, forward atmosphere models and retrieved atmosphere models.

\section{Existing results for WASP-17b and HD\,149026b}
\label{sec:existing_results}

Of the planets in our sample, two have published JWST observations to date. \cite{Grant2023} presented a MIRI/LRS spectrum of WASP-17b from 5--12\,\micron, combined with previously published HST and Spitzer observations of the planet \citep{Alderson2022_w17}. While the different atmosphere modelling procedures they used resulted in a range of metallicities and C/O ratios, their result pointed towards a depleted H$_2$O abundance and a super-solar C/O. \cite{Grant2023} interpreted this as due to the formation of high temperature aerosols depleting the O from the observable atmosphere. Since WASP-17b is a misaligned planet, its super-solar C/O and sub-solar O/H could also be a natural outcome of it not accreting O-rich material from the inner disc (\citetalias{PenzlinBooth2024} \citeyear{PenzlinBooth2024}) as it likely underwent high-eccentricity migration due to its high obliquity (Table \ref{tab:sample}). 

\cite{Bean2023} observed an emission spectrum with JWST/NIRCam of the aligned hot Jupiter HD\,149026b, which revealed a metal-rich atmosphere ($59-276\times$ solar) and a super-solar C/O ($0.84 \pm 0.03$). However, in a re-analysis, \cite{Gagnebin2024} showed that the planet's emission spectrum could also be fit with a lower, yet still super-solar, metallicity of $20^{+11}_{-8}\times$ solar and super-solar C/O of $0.67^{+0.06}_{-0.27}$ by using self-consistent 1D radiative-convective-thermochemical equilibrium models. Both conclusions are qualitatively consistent with the silicate rainout disc model (Figure \ref{fig:anna_models} and \citetalias{PenzlinBooth2024} \citeyear{PenzlinBooth2024}). HD\,149026b is the only planet in our sample that will not be observed with NIRSpec/G395H in transmission, with the published data coming from NIRCam/F322W2+F444W emission observations \citep{Bean2023}, which cover a similar wavelength range to NIRSpec/G395H (2.3--5.0\,\micron). We will perform a reanalysis of these data using our own reduction tools (Section \ref{sec:data_reduction}) to ensure uniformity and investigate the implications on the measured C/O distribution for the aligned planets when including and excluding this data set. 

While it is tempting to begin to draw conclusions from the published observations of WASP-17b and HD\,149026b, we stress the importance of waiting for the combined, homogeneously observed and analysed sample when trying to understand planet formation. 

\section{Additional science enabled by our homogeneous sample}
\label{sec:ancillary_science}

Given that our observing programme will produce eight homogeneously generated and analysed, high signal-to-noise JWST spectra of hot Jupiters, there are multiple additional science questions that can be explored, beyond our primary goal of determining whether C/O and Z depend on migration in a measurable way. For example, we will also investigate whether Z depends on planetary mass. In the solar system there is an inverse relationship between mass and metallicity among the gas giants \citep[e.g.,][]{Kreidberg2014}. Determining whether such a trend exists among exoplanets has long been a goal of observations \citep[e.g.,][]{Kreidberg2014,Wakeford2018,Welbanks2019}. Our sample of eight planets span Saturn to super Jupiter masses ($\sim 0.4-1.3$\,\Mjup), allowing for informative comparisons with the solar system trend. 

In addition to the 1D forward and retrieval modelling we will perform for our sample, we will investigate the atmospheres of our exoplanets in 3D. Since all of our targets are on short orbital periods they are expected to be tidally locked. This leads to a large day-to-night temperature gradient between their permanent daysides and permanent nightsides which, in turn, drives a super-rotating equatorial jet that carries hot dayside gas to the nightside and cool nightside gas to the dayside \citep[e.g.,][]{Showman2010}. The result is hotter evening limbs and cooler morning limbs, and hence different gas phase chemistry and cloud coverage at each limb. Using a combination of ingress/egress light curve fits and light curve fits with two planetary radii for the morning and evening limb, \cite{Espinoza2024} have measured limb differences in a planet's transmission spectrum with JWST. We will apply the same techniques, using \texttt{Tiberius} and \texttt{catwoman}\footnote{\url{https://github.com/KathrynJones1/catwoman}} \citep{Jones2020} to search for limb asymmetries in our data.

Furthermore, we will investigate how the F star hosts of our sample drive photochemistry in their planets' atmospheres. Given the detection of the photochemical product SO$_2$ in the transmission spectrum of WASP-39b \citep{Alderson2023,Rustamkulov2023,Tsai2023}, a Saturn-mass giant around a mid-G star, it is plausible that we will also see SO$_2$ in our planets. If we see evidence for SO$_2$ in a planet's spectrum, this will necessitate photochemical modelling for that planet within the primary science programme. Our inferences of sulphur abundances will allow for additional precise constraints on atmospheric metallicity \citep[e.g.,][]{Tsai2023} and will motivate parallel investigations into if and how a planet's history can be inferred from its sulphur abundance.

Our investigations of 3D effects and photochemistry will be backed up by a comparison of each planet's observed spectrum to the spectrum predicted by a 3D climate model (also known as a general circulation model, or GCM) of a hot Jupiter atmosphere. This comparison will reveal how the interactions between atmospheric circulation, radiative transfer, disequilibrium thermochemistry \citep[e.g.,][]{Zamyatina24_quenchingdriven} and photochemistry, in 3D, impact the exoplanets' morning and evening limb spectra. Furthermore, this will allow us to determine to what extent the limb-averaged C/O and Z we measure from our transit spectra are representative of the overall atmosphere composition. Among the GCM outputs will be longitude-latitude maps of molecular abundances as a function of pressure and temperature, which we will use to inform our 1D models. The BOWIE-ALIGN programme sample could also serve as a high-quality observational reference for model intercomparison projects, such as those within the CUISINES framework \citep{Fauchez21_workshop,Sohl2024}. The homogeneity of data would allow to benchmark single-column models and 3D GCMs in terms of e.g.\ their simulated transmission spectra, for aligned and misaligned hot Jupiters. This will help to validate our theoretical predictions and make the conclusions of the programme more robust.

Separately, we will perform cross-correlation analyses for all of the planets in our sample to search for isotopologues in the planets' atmospheres. As \cite{Esparza-Borges2023} showed, one-pixel resolution spectra from NIRSpec/G395H can be used to detect molecules via cross-correlation spectroscopy, in a similar way to that done using high-resolution ground-based instrumentation \citep[e.g.,][]{Snellen2010,Rodler2012,Brogi2016}. In the case of WASP-39b, \cite{Esparza-Borges2023} showed that the signal-to-noise in the cross-correlation function was maximised by the inclusion of CO isotopologues, demonstrating the sensitivity of JWST to isotopologues. This will provide additional information regarding our samples' formation since elemental isotopes and molecular isotopologues may be related to formation history \citep[e.g.,][]{Clayton2004}.

Finally, our JWST observations will allow us to measure extremely precise flux-calibrated 2.8--5.2\,\micron\ spectra of F stars from around 6300 to 6700\,K. These spectra will enable robust tests of 3D stellar magnetohydrodynamics (MHD) atmospheric models, specifically those from the \texttt{MuRAM} code \citep[]{Vogler2003,Vogler2005}.

We will make our data products publicly available after each planets' analysis, which will enable the community to pursue additional ancillary science programmes.

\section{Summary}
\label{sec:summary}

We are undertaking a survey with JWST to test whether an exoplanet's atmospheric C/O and metallicity can be observationally linked to its migration history. Specifically, we will observe the near-infrared spectra of a sample of eight hot Jupiters, four of which are aligned with the stellar spin axis and four of which are misaligned. Crucially for our test, all of our targeted exoplanets orbit F stars above the Kraft break, where tidal realignment is expected to be inefficient, meaning that their obliquities retain information about their migration histories. 

\citetalias{PenzlinBooth2024} \citeyear{PenzlinBooth2024} explored the range of C/O and metallicity that can result from a consideration of multiple different protoplanetary disc and accretion parameters specifically for planets formed around F stars. The principal observationally-testable results from their modelling effort are:

\begin{enumerate}
    \item $\mathrm{C/O}_{\mathrm{aligned}} < \mathrm{C/O}_{\mathrm{misaligned}}$: this can be explained by the late accretion of solid material (planetesimals) by aligned planets during their migration. This would be direct evidence that accretion from the inner disc is important in setting a planet's C/O and that the aligned planets and misaligned planets did indeed undergo different migration pathways. \\
    \item $\mathrm{C/O}_{\mathrm{aligned}} > \mathrm{C/O}_{\mathrm{misaligned}}$: this scenario is possible if, during the late accretion of O-bearing silicates in the inner disc, the silicates do not evaporate their oxygen upon accretion. In this case, the atmospheric composition will have been dominated by accreted gas, volatiles and carbon-rich refractories. Like result (i), this would be direct evidence for the influence of migration on atmospheric composition. \\
    \item $\mathrm{C/O}_{\mathrm{aligned}} \approx \mathrm{C/O}_{\mathrm{misaligned}}$: this result would indicate either that the majority of hot Jupiter assembly is completed before migration or that obliquity is not a sensitive tracer of migration for planets around F-type stars.
\end{enumerate}

In this paper, we have described our observational test of these predictions, including the target selection and sample size, which we investigated using a combination of 256 1D atmospheric forward models, JWST noise simulations and atmospheric retrievals. 

Our simulated transmission spectra show we will measure the predicted difference in C/O of aligned versus misaligned planets with confidences between 0 and $>3\sigma$, depending on where our targeted exoplanets sit in C/O--[O/H] parameter space (Figure \ref{fig:anna_models}). We will be sensitive to result (i) if the atmospheric metallicities approach $\sim 30\times$ solar, while the greater differences in C/O between the two populations for result (ii) means we will be sensitive to these over a wide range of metallicities. If we measure either result (i) or (ii), this would represent the first observational confirmation that atmospheric composition traces formation environment. If we observe super-solar metallicities, this would indicate the accretion of solid material, like the solar system, while sub-solar metallicities would suggest that hot Jupiter assembly differs from the giant planets in the solar system. 

A caveat to our experiment is the assumption that disc-migration results in aligned systems while disc-free migration results in misaligned systems. Other scenarios, like coplanar high-eccentricity migration \citep{Petrovich2015}, could contaminate this distinction. If we measure a negligible difference between the compositions of aligned and misaligned planets, result (iii), this could be evidence for a common migration history and thus that obliquities are not a sensitive tracer of migration, even around F stars.

Ultimately, our comparative survey of a population of carefully-selected hot Jupiters will provide a rich dataset with which to test the dependence of atmospheric composition on formation and migration.

\section*{Acknowledgements}

J.K. acknowledges financial support from Imperial College London through an Imperial College Research Fellowship grant. J.E.O. and R.A.B are supported by Royal Society University Research Fellowships and A.B.T.P. received support from their Enhancement Awards. This project has received funding from the European Research Council (ERC) under the European Union’s Horizon 2020 Framework Programme (grant agreement no. 853022, PEVAP).  N.J.M., D.E.S. and M.Z. acknowledge support from a UKRI Future Leaders Fellowship [Grant MR/T040866/1], a Science and Technology Facilities Council Nucleus Award [Grant 954 ST/T000082/1], and the Leverhulme Trust through a research project grant [RPG-2020-82]. P.J.W. acknowledges support from the UK Science and Technology Facilities Council (STFC) through consolidated grants ST/T000406/1 and ST/X001121/1. V.P. acknowledges support from the UKRI Future Leaders Fellowship grant MR/S035214/1 and UKRI Science and Technology Facilities Council (STFC) through the consolidated grant ST/X001121/1. C.E.F. acknowledges support from the Swiss National Science Foundation (SNSF) Mobility Fellowship [grant no. P500PT\_203110] and the ERC under the European Union’s Horizon 2020 research and innovation program [grant agreement no 805445]. E.A. would like to thank C. McDonald and D. Veras for insightful discussions.

\section*{Data Availability}

The \texttt{PLATON} atmosphere models, retrieval results and \texttt{PandExo} spectral uncertainties are available on Zenodo at this link: \url{https://zenodo.org/records/13902706}. 




\bibliographystyle{mnras}
\bibliography{main} 




\appendix

\section{Posteriors for cloudy models}
\label{sec:apx_cloud}

This appendix includes the retrieved $\log Z$ and C/O posteriors when including a grey cloud deck at 1\,mbar.

\begin{figure*}
    \centering
    \includegraphics[scale=0.8]{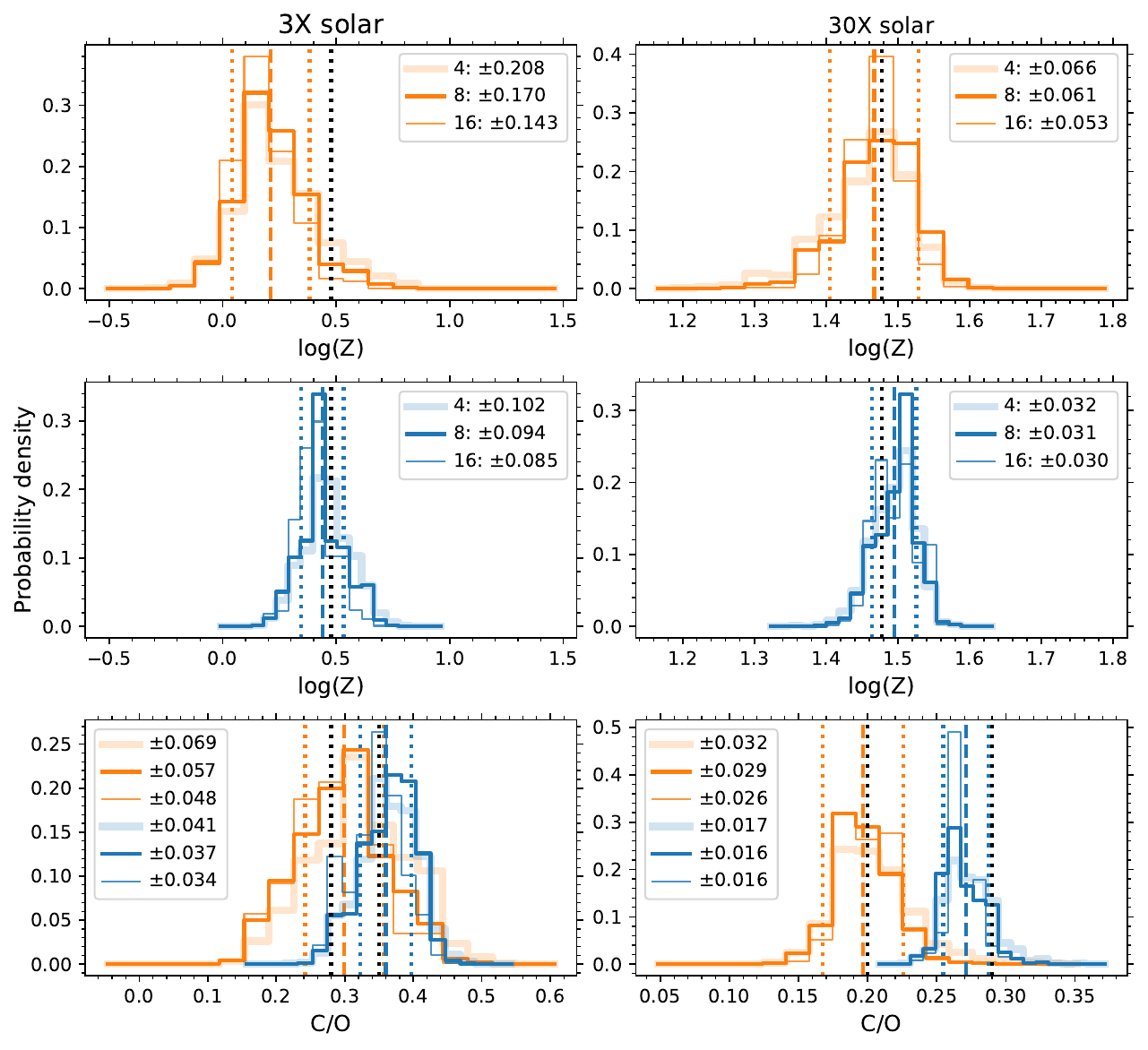}
    \caption{The retrieved $\log Z$ and C/O posteriors from the model atmospheres with $\log Z$ and C/O set by the fiducial disc model (scenario 1), with the addition of a grey cloud deck at 1\,mbar. See Figure \protect\ref{fig:posteriors_fiducial} for a description of the axes.}
    \label{fig:posteriors_fiducial_mbar_cloud}
\end{figure*}

\begin{figure*}
    \centering
    \includegraphics[scale=0.8]{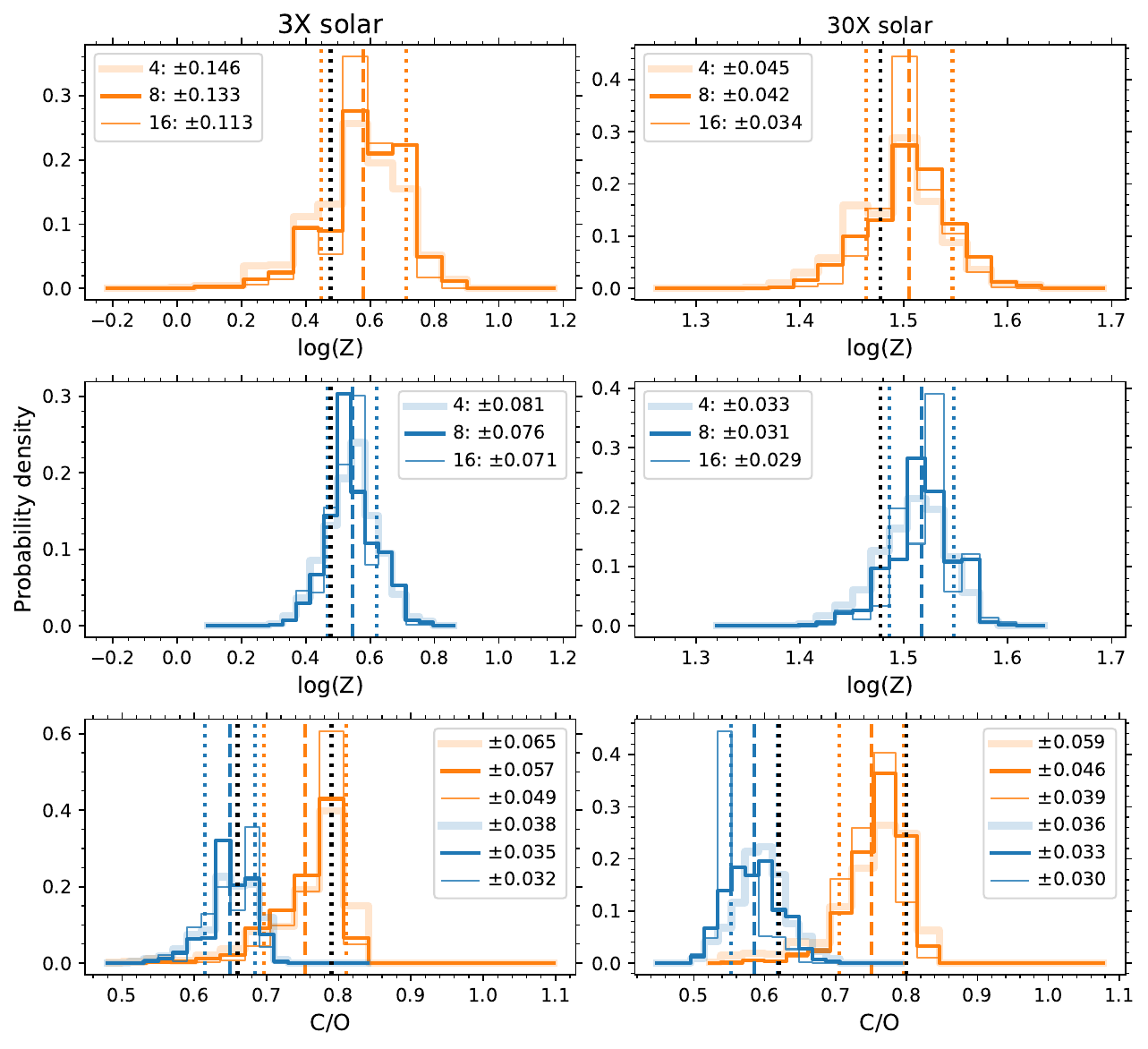}
    \caption{The retrieved $\log Z$ and C/O posteriors from the model atmospheres with $\log Z$ and C/O set by the silicate rainout disc model (scenario 2), with the addition of a grey cloud deck at 1\,mbar. See Figure \protect\ref{fig:posteriors_fiducial} for a description of the axes.}
    \label{fig:posteriors_noSi_mbar_cloud}
\end{figure*}


\bsp	
\label{lastpage}
\end{document}